# Disordered nano-wrinkle substrates for inducing crystallization over a wide range of concentration of protein and precipitant

By Anindita Sengupta Ghatak[1] and Animangsu Ghatak[1,2,*]

[1] Department of Chemical Engineering and DST Unit on Soft Nanofabrication, Indian Institute of Technology, Kanpur, 208016 (India)

[2] Max Plank Institute for polymer research, Ackermannweg 10, D-55128, Mainz, Germany

[*] Prof. A. Ghatak Corresponding-Author, Author-Two, E-mail: aghatak@iitk.ac.in

## ABSTRACT

There are large number of proteins, the existence of which are known but not their crystal structure, because of difficulty in finding the exact condition for their crystallization. Heterogeneous nucleation on disordered porous substrates with small yet large distribution of pores is considered a panacea for this problem, but a universal nucleant, suitable for crystallizing large variety of proteins does not really exist. To this end, we report here a nano-wrinkled substrate which displays remarkable ability and control over protein crystallization. Experiments with different proteins show that on these substrates, crystals nucleate even at very low protein concentration in buffer. Small number of very large crystals appear for precipitant concentrations varied over orders of magnitude ~0.003–0.3M; for some proteins, crystals appear even without addition of any precipitant, not seen with any other heterogeneous substrates. In essence, these substrates significantly diminish the influence of the above two parameters, thought to be key factor for crystallization, signifying that this advantage can be exploited for finding out crystallization condition for other yet-to-be-crystallized proteins.

# 1. INTRODUCTION

Crystallization of protein molecules from supersaturated solution is a complex phenomenon, dictated by several variables: rate of evaporation of water, concentration of proteins, chemical nature of precipitants and additives, temperature of the surrounding environment, pH of the buffer solution and even physical/chemical heterogeneity of the substrate. As a result, determining the exact recipe and range of values of different parameters for generating diffraction quality crystals of protein poses a significant scientific challenge. Therefore, despite many attempts, still there are several proteins which have not yet been crystallized with size suitable for X-ray diffraction. Naturally, the search for a suitable process/material, which can decrease the potential energy barrier for formation of the critical nuclei, has been a subject of vigorous experimental and theoretical research[1–3], leading to the realization that a chemically and/or physically modified heterogeneous substrate[4–8] can possibly serve as a universal nucleant[9]. An example of such substrate is porous material with pore sizes approximately same as that of a critical nucleus; for example, porous silicon and glass surfaces[10–13], porous polymeric membrane have pores of sizes 40-200 nm[14], bioactive gel-glass[5], nano-porous materials[15,16] with pore sizes in the range of 2−10 nm are indeed found to be effective nucleating agent. A direct correlation exists between the pore size $d_p$ and the radius of gyration $R_g$ of the protein molecules: while pores with $d_p < R_g$ do not contribute to nucleation, those with $d_p > R_g$ do not diminish the energy barrier sufficiently enough for nucleation to occur. Furthermore, uniformity in pore size too is detrimental as it does not facilitate breaking of symmetry and induce nucleation. Uniformity in diameter of nucleant also allows only protein molecules of specific size to be crystallized[15]. Therefore, it is hypothesized that disordered porous materials are most suitable for promoting nucleation[5,10] thus underlining the need of a substrate with very small pores but with wide distribution in pore size. It is in this context that we present here a novel nano-wrinkled substrate, the local radius of curvature of which varies over a wide range from -10nm to 10nm and over a length scale of 125nm.

These substrates are prepared in a two stage process. In the 1$^{st}$ stage, a thin strip of an elastomer, e.g. poly(dimethylsiloxane) (PDMS) is stretched to a desired extension ratio, λ ~ 1.2 – 1.4 and is plasma oxidized to generate a thin silica crust and is then released instantaneously leading to longitudinal surface folds, because of the difference in deformability of the stiff skin and the soft under-layer[17-19]. The atomic force microscopy (AFM) images in figure 1a and b depict these 1$^{st}$ generation wrinkles (hereon referred to as "$S_1$" surface) which appear far from sinusoidal in contrast to similar experiments but with slower release of slightly plasma-oxidized films[17]. The wrinkles remain asymmetric with cylindrical folds merging along a deep furrow forming a sulcus like structure; these are also the locations where the curvature of these surfaces becomes maximum. The maximum curvature $\kappa_{max}$ however does not exceed the inverse material length scale $(\gamma/\mu)^{-1}$, because, surface tension does not allow larger curvature to be achieved[20]. For PDMS with surface tension, γ = 22 mN/m and shear modulus μ = 0.6 MPa, $\kappa_{max}$ is found to be ~0.02 nm$^{-1}$, thus defining a limiting lengthscale for most soft-lithography methods. Increasing the modulus of the elastomer does not help increasing the $\kappa_{max}$, because then the advantage of using a soft, deformable material for patterning is compromised. We overcome this difficulty by choosing a material which is of low modulus at the bulk, but stiff at the vicinity of its surface. In particular, wrinkled substrates prepared in the 1$^{st}$ stage (Figure 1a–b) is used for this purpose, because, the presence of the thin crust increases its modulus close to the surface without altering it at the bulk. When we subject such a substrate to a second cycle of above process, finer 2$^{nd}$ generation wrinkles as represented by Figure 1c–d (hereon referred to as "$S_2$" surface) with $\kappa$ as large as 0.1 nm$^{-1}$ appear at the location of merger of folds. In addition, the standard deviation of the distribution of curvature too remains very large. Importantly, curvature varies continuously from positive to the negative maxima, both of which are useful for crystal nucleation[4], an advantage not available with conventional porous substrates, for which, the surface remains essentially flat except at the location of pores, where it remains large but nearly equal[13] for all pores.

## 2. EXPERIMENTAL SECTION

**2.1. Materials.** Ferritin, a protein extracted from horse spleen and thaumatin extracted from bacteria thaoumatococcus daniellii were purchased from Sigma chemicals. Glucose isomerase obtained from bacteria streptomyces rubiginosus and xylanase from trichoderma reesei were purchased from Hampton research. ADA buffer and Na/K tartrate were purchased from Hampton research. CdSO$_4$.8H$_2$O, Sodium azide, CaCl$_2$.2H$_2$O, NaCl, Tris HCl, PEG 8000, (NH$_4$)$_2$SO$_4$ were all purchased from Marc. Izit dye was purchased from Hampton Reseach. The proteins were used for the experiments without further purification. Deionized water obtained from Mili Q system was used throughout the study. Sylgard 184 elastomer was obtained from Dow Corning and octadecyltrichlorosilane (OTS) molecules were obtained from Sigma chemicals.

**2.2. Method of preparation of protein solutions.**

**2.2.1. Ferritin.** The protein solution of 50mg/ml was obtained from the supplier, which was diluted to 20mg/ml with deionized water. In addition, 10% 3CdSO$_4$. 8H$_2$O, 3 mM NaN$_3$ and (NH$_4$)$_2$SO$_4$ in the concentration range of 0.0 – 2.0 M were used for the crystallization experiments. Each drop of the pre-crystallization solution was prepared by mixing 0.8 μl of each of the above CdSO$_4$. 8H$_2$O, NaN$_3$ and (NH$_4$)$_2$SO$_4$ solution and 2.4 µl of protein solution. Thus the protein concentration in the resultant pre-crystallization solution was 10 mg/ml whereas the concentration of (NH$_4$)$_2$SO$_4$ lie in the range of 0.0 – 0.3 M.

**2.2.2. Thaumatin**. The protein was obtained from Sigma chemicals as dry flakes, which was dissolved in 0.1M ADA buffer (pH = 6.5) to attain a concentration of 30 mg/ml. Na/K tartrate in concentration range: of 0.0–1.5 M was used as the precipitant for crystallizing the protein. 2 µl of each of the protein solution and that of Na/K tartrate was mixed in the pre-crystallization solution. Thus the resultant solution had a protein concentration of 15 mg/ml and the precipitant concentration in the range of 0.0 – 0.75 M.

**2.2.3. Glucose Isomearse**. The protein solution obtained at 33mg/ml was diluted to 10mg/ml concentration with deionized water. 0.1M HEPES (pH = 7.5) and (NH$_4$)$_2$SO$_4$ in the concentration range of 0.2 – 2.0M were used as precipitants for crystallization. Each drop of the pre-crystallization solution contained 1.0 µl of each of the HEPES, (NH$_4$)$_2$SO$_4$ solution and 2 µl of the protein solution. The resultant solution thus had 5 mg/ml of protein and 0.05 – 0.5 M of (NH$_4$)$_2$SO$_4$.

**2.2.4. Xylanase**. The protein solution obtained at 36mg/ml was diluted to 10 mg/ml with deionized water. For the crystallization, we used 0.1M NaCl, 0.05M Tris HCl (pH = 8.5), 15% (w/v) PEG 8000, and CaCl$_2$ in the concentration range of (0.2–1.2) M. Each drop contained 0.7 µl of each of NaCl, Tris HCl, PEG 8000, CaCl$_2$ 2H$_2$O solution and 2.8 µl of the protein solution. Thus the protein and precipitant concentration in the resultant pre-crystallization solution was maintained at 5 mg/ml and 0.0025 – 0.15 M respectively.

**2.3. Method of preparation of nano-wrinkled surfaces.**

Thin strips of PDMS film (length 25 mm, width 10 mm, thickness ~ 350μm) were uniaxially stretched to a desired extension ratio using a home-made gadget, following which, the films were exposed to radio frequency oxygen plasma (Harrick Plasma, model PDC-32G, pressure ~0.05 Torr, power ~6.8W, time ~ 4 min). The stretched films were then released instantaneously without using any motion control device. This process led to the appearance of surface wrinkles because of difference in stiffness of the thin silicate crust (elastic modulus ~76 GPa) and the elastic substrate (modulus ~3 MPa)[19]. Wrinkles of different periodicity and amplitude were generated by varying the extension ratio λ ~ 1.2 – 1.4. Here-on we refer to these surfaces with 1st generation wrinkles as the "$S_1$" surface. The PDMS films patterned with the 1st generation wrinkles were subjected to a repeated sequence of the above three steps (time for plasma oxidation ~ 1 min), which yielded the 2nd generation wrinkles. We refer to these surfaces as "$S_2^"$" surface. During preparation of both the 1st and 2nd generation samples, the PDMS strip was stretched uniaxially along identical direction and to identical extension ratio. In addition to these two different types of substrates, we used also flat wrinkle free surface of PDMS which we turned hydrophilic by small degree of plasma oxidization (pressure ~0.05 Torr, power ~6.8W, time ~ 1 min). We refer to these surfaces as "S" which we used as the control surface.

**2.4. Atomic Force Microscopy (AFM) measurement**. The wrinkled substrates were imaged in tapping mode by using Pico Plus Atomic force microscope, (Molecular Imaging) at room temperature. The scan size was 2μm x 2μm and 5μm x 5μm.

**2.5. Method for obtaining curvature from the AFM images of nano-wrinkled substrates.** In order to calculate the curvature, such a profile was first divided into several segments, each of which was fitted to a polynomial equation: $y = \sum_{n=0 \to n} a_0 x^n$. After several trials, a 10th order polynomial was found to perfectly capture the surface profiles, including the junction of the folds where curvature was maximum. The curvature was then calculated as $\kappa = d^2y/dx^2$ and plotted in figure 2(b). Because of uni-axial extension, the curvature along $z$ is expected to be negligible relative to that along $x$, except at the vicinity of the defects, therefore was not considered for calculation.

**2.6. Method of carrying out x-ray diffraction analysis.** X-ray diffraction data were collected from the crystals by using Rigaku MicroMax007HF X-ray source with a copper rotating-anode generator equipped with Varimax optics, a MAR345dtb image-plate detector and an Oxford Cryosystem 700 series cryostream.

## 3. RESULTS AND DISCUSSION

### 3.1. Energy barrier for nucleation on nano-wrinkled substrates.

An estimate of rate of nucleation on these substrates can be obtained by considering that the free energy barrier $\Delta F^*$ for nucleation of crystals is a function of curvature $\kappa$ at any location, so that the rate of nucleation can be expressed as[5,21]: $R_\kappa = \nu \cdot \exp(-\Delta F^*(\kappa)/k_B T)$. Here, $\nu$ is the rate (time$^{-1}$) at which material gets transported to the nucleation site and $k_B$ and $T$ are respectively the Boltzmann constant and the absolute temperature. At a location infinitesimally away from where it is maximum, the curvature can be expressed by Taylor series expansion as: $\kappa = \kappa_{max} + (\partial \kappa / \partial x)\big|_{x=x_{max}} \Delta x \approx \kappa_{max} + \kappa'\big|_{x=x_{max}} \eta$, where, $\kappa'\big|_{x=x_{max}}$ is the gradient of curvature and $\Delta x$ is a characteristic length-scale which can be represented by the wavelength $\eta$ of the wrinkles. These two parameters intrinsically accounts for the local variation in surface topography, in contrast to an equivalent porous substrate, for which the asphericity in the shape of pores are required to be accounted via a shape factor, an empirical quantity[5]. The energy barrier is expected to be minimum at the vicinity of maximum curvature $\kappa_{max}$ but increase away from it according to the following expansion about the minimum energy barrier: $\Delta F^*(\kappa) = \Delta F^*(\kappa_{max}) + D(\kappa_{max} - \kappa)^2/2$, where $D$ is the stiffness of the energy landscape. Combining the above two expressions, we obtain a relation for the energy barrier in terms of the gradient in curvature: $\Delta F^*(\kappa) = \Delta F^*(\kappa_{max}) + D(\kappa'|_{x=x_{max}})^2 \eta^2/2$. If $p(\kappa)$ is the probability density function of occurrence of curvature, $\kappa$ then the rate of nucleation of a crystal can be estimated as $\nu \int_{-\infty}^{\infty} \exp\left[-\frac{\Delta F^*(\kappa)}{k_B T}\right] \kappa p(\kappa) d\kappa$. An expression for the rate of nucleation is then obtained by considering normal distribution for $p(\kappa)$ with $\kappa_\mu$ as the mean curvature and $\sigma_0$ as the standard deviation:

$$R = \nu \int_{-\infty}^{\infty} \exp\left[-\frac{\Delta F^*(\kappa)}{k_B T}\right] \kappa p(\kappa) d\kappa = \nu \exp\left[-\frac{\Delta F^*(\kappa_{max})}{k_B T} - \frac{D(\kappa'|_{x=x_{max}})^2 \eta^2}{2 k_B T} - \frac{(\kappa_{max} - \kappa_\mu)^2}{2\sigma_0^2}\right]$$

The above relation suggests that the rate of heterogeneous nucleation is a function of several topographical parameters of the substrate. For example, the mean curvature $\kappa_\mu$ and the distribution in curvature $\sigma_0$ are important, as increase in both these quantities, enhances the nucleation rate $R$; in addition, gradient in curvature $\kappa'$ too is important as $R$ is expected to increase with decrease in $\kappa'$, an effect not observable with porous substrates. Furthermore, since the curvature of wrinkled substrates varies from positive and negative, the above relation suggests that the crystals are expected to nucleate faster on a concave surface for which the curvature turns negative, a conclusion also arrived from the route of molecular dynamic simulations[4] of crystallization of colloidal crystals. In particular, spatial variation in curvature over a large range allows protein molecules to self-select a curvature during aggregation to form a cluster.

**3.2. Curvature of wrinkles.**

We then obtain the numerical estimates of the distribution of curvature and the spread of these distributions for various substrates used in our experiments by analyzing their corresponding images using the image processing software, ImageJ. We first obtain from the AFM image, the height profile of a wrinkled surface along a straight horizontal line drawn along $x$ at any random location along $z$. The plot in Figure 1e shows a typical profile obtained for a 2$^{nd}$ generation wrinkle prepared with $\lambda = 1.4$ (referred by S$_2$|$_{40\%}$). The profile of curvature (Figure 1f) appears oscillatory with several positive and negative peaks. The number of peaks appears more than the number of folds, because of roughness of the surface over length-scales much smaller than that of the wavelength of the wrinkles and the noise associated with AFM imaging and the subsequent measurement. In order to have realistic estimation of the distribution of curvature, we filter out the peak curvatures having absolute value smaller than a threshold limit, only ones exceeding this limit are considered for further calculation. For any substrate, the threshold is set to be 20% of $\kappa_{max}$ attained, e.g., for the S$_2$|$_{40\%}$ surface, this limit is found to be 0.025 nm$^{-1}$. Such a set of peak curvatures compiled after analyzing 150 randomly chosen line-profiles show that curvature for this surface varies from 0.027 nm$^{-1}$ to 0.21 nm$^{-1}$. Figure 1g shows probability distribution of curvature of several such wrinkled surfaces. For all cases, with increase in curvature from the threshold value, the probability of its occurrence first increases, reaches a maxima and then decreases to a small value for very large curvature. The curvature at which this maxima occurs is considered as the mean curvature, $\kappa_0$ and the standard deviation of the distribution $\sigma_0$ is then calculated by considering it to be a normal distribution. The maximum and mean curvatures and the standard deviation, all differ for different surfaces and vary distinctly for 1st and 2nd generation wrinkles (Figure 1h). For example, for the $S_2|_{40\%}$ surface, the probability of $\kappa \geq 0.1 \text{nm}^{-1}$ is found to be @ 0.4%, whereas for a $S_2|_{30\%}$ surface, the maximum curvature, $\kappa_{max}$ is found to be 0.096 nm$^{-1}$. For $S_1|_{30\%}$ and $S_1|_{40\%}$ surfaces, $\kappa_{max}$ is found to be 0.042 nm$^{-1}$ and 0.05 nm$^{-1}$ respectively. Apart from $\kappa_{max}$ being large, the other important parameter is $\sigma_0$ which for the $S_2|_{40\%}$ and $S_2|_{30\%}$ surfaces are found to be 0.017 and 0.0144 nm$^{-1}$ respectively. These values are significantly larger than the corresponding $S_1$ wrinkles, $\sigma_0 = 0.0062$ and 0.0055 nm$^{-1}$. For the $S_2|_{20\%}$ wrinkles, all these parameters were found to be smaller by about an order of magnitude than the other cases.

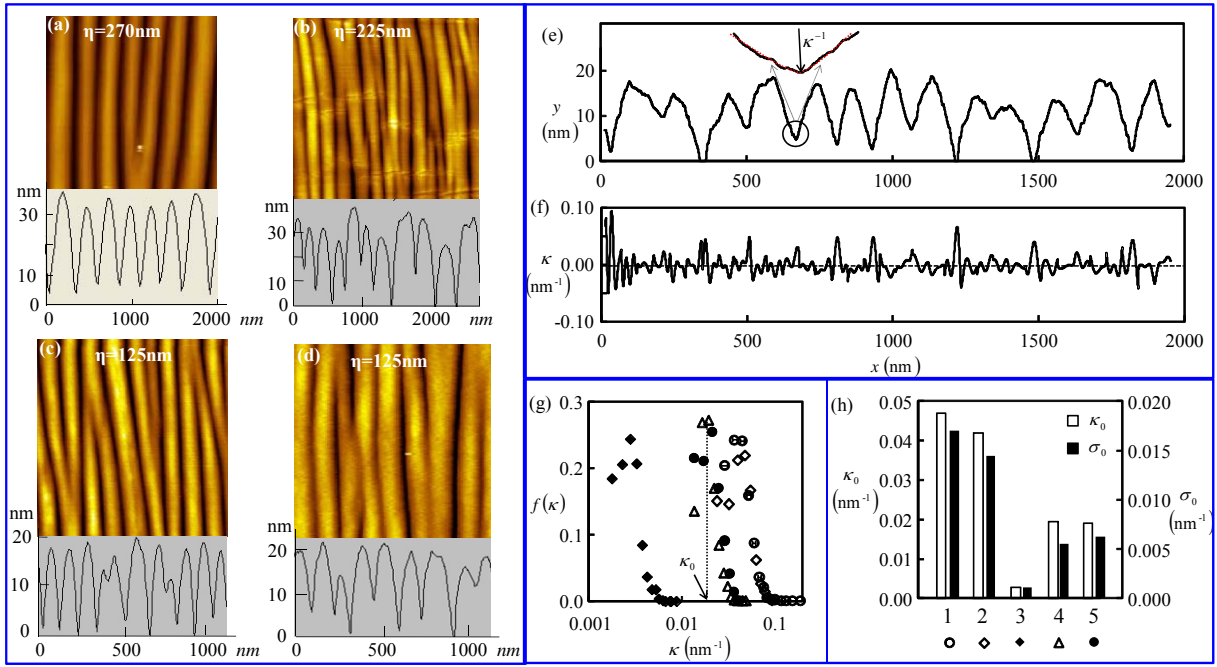

**Figure 1: (a − d) Atomic force microscopy (AFM) images of wrinkle patterns generated on PDMS films. Images (a) and (c) represent respectively the 1st and 2nd generation wrinkles obtained by stretching the PDMS film to extension ratio: $\lambda = 1.3$. Similarly images (b) and (d) represent the same for $\lambda = 1.4$. (e) The AFM images a-d were used for obtaining typical profiles of the surface of a wrinkled film along a horizontal line drawn perpendicular to the**

longitudinal axes of the wrinkles. The specific surface corresponding to the data in plot (e) was prepared via double extension of 40%-40% (represented by $S_2|_{40\%}$). The inset depicts that small segment of the profile was fitted to a polynomial curve, which was then used for obtaining the curvature. (f) Several such small segments were used to obtain the curvature $\kappa$ as a function of spatial distance $x$. Several line-profiles were chosen randomly from the AFM image for obtaining the corresponding curvatures which were found to be oscillatory consisting of several positive and negative peaks. The positive peaks larger than a threshold value was considered and the probability of occurrence of such a peak curvature was plotted in figure (g). The semi-log plot in Figure (g) shows several such distribution functions of positive-peak curvatures for different wrinkled surfaces. Symbols ○, ◇ and ◆ represent wrinkled surfaces $S_2|_{40\%}$, $S_2|_{30\%}$ and $S_2|_{20\%}$ respectively; symbols △ and ● represent respectively $S_1|_{40\%}$ and $S_1|_{30\%}$. These distribution functions were assumed to be normal and the corresponding mean $\kappa_\mu$ and standard deviation $\sigma_0$ were obtained. The bar chart (h) depicts these values for different surfaces considered in plot (g).

### 3.3. Schematic of experimental set-up.

In order to determine, how wrinkle patterns of different $\kappa_0$ and $\sigma_0$ affect nucleation of protein crystals, these substrates were used in the crystallization set up[8] of Figure 2 consisting of two rigid plates placed parallel to each other with a small gap ~120 μm maintained by two spacers. Here the top plate was silanized by coating it with self-assembled monolayer (SAM) of octadecyltrichlorosilane (OTS) molecules, which renders this surface hydrophobic, while the nano-wrinkled poly-(dimethylsiloxane) (PDMS) film attached to a rigid glass slide was used as the bottom surface, which was hydrophilic. A small volume of 4μl of the protein-buffer-precipitant solution was placed between these two substrates to form a liquid disk of diameter ~2.0 mm. The whole set up was placed inside the controlled environment of a constant temperature bath maintained at 20ºC. In contrast to conventional hanging drop method, this set-up has the advantage that it allows controlled evaporation of water via fine adjustment of the gap between the film and the plate. The resultant crystals are amenable for easy harvesting for diffraction analysis or further downstream processes. Since, several liquid disks can be formed side by side, the experiment allows also for easy scale up of screening of the crystallization condition.

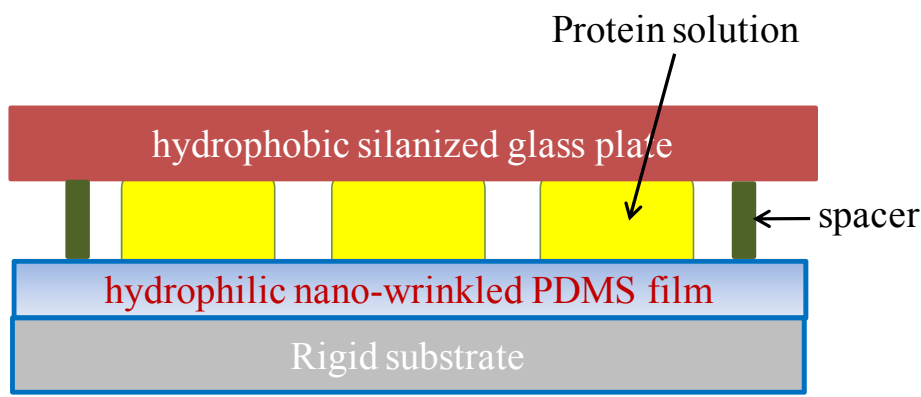

**Figure 2. Schematic of the crystallization set up with parallel plate geometry**

### 3.4. Crystallization of Ferritin.

Depending upon the topography of the patterned PDMS film, crystallization was observed to proceed differently for the protein molecules used in our experiments. Figure 3a–l show the results for iron storing protein ferritin (molecular weight = 474kDa) crystallized on flat, wrinkle free substrate S and on surfaces with 1$^{st}$ and 2$^{nd}$ generation wrinkles, i.e. $S_1|_{40\%}$ and $S_2|_{40\%}$ respectively. In conventional hanging drop method[22], ferritin has usually been crystallized using cadmium sulfate (80mM $3CdSO_4 \cdot 8H_2O$), sodium azide (3mM) and ammonium sulphate (1.0M) as the precipitants. In our experiment, we varied concentration of $(NH_4)_2SO_4$ from 0.0 – 0.3 M, keeping that of other species in the buffer constant (10mg/ml ferritin, 1.6% $3CdSO_4 \cdot 8H_2O$ and 0.5mM $NaN_3$). Negligible nucleation was observed on smooth, featureless surface S when the concentration of ammonium sulfate, $C_{ammonium\_sulfphate}$ was smaller than 0.003 M (Figure 3a); beyond this limit, small crystals of ferritin appeared within twelve hours, which grew to their maximum size within 2 days (Figure 3b). Large number of crystals appeared and grew with longest dimension, $l$ finally reaching 30.2 ± 3.2 µm for $C_{ammonium\_sulphate}$ ~0.03M (Figure 3c). The crystals were even larger with $l$ ~ 63.2 ± 20 µm and 84.8 ± 6.8 µm for $(NH_4)_2SO_4$ concentration 0.16 M and 0.3 M respectively (Figure 3d). On $S_1|_{40\%}$ surface with wrinkles of intermediate curvature, crystals were larger, with longest dimension attaining $l$ ~ 150.4 ± 3.4 µm for $C_{ammonium\_sulfate}$ = 0.3 M (Figure 3h). The crystal size decreased with decrease in $C_{ammonium\_sulfate}$, finally getting negligibly small for 0.0M. In contrast to these surfaces, crystal sizes were significantly larger on the $S_2|_{40\%}$ surface (Figure 3i–l). For example, crystals of size $l$ ~ 55.3 ± 12.3 µm appeared even when no $(NH_4)_2SO_4$ was used[15]; large number of tiny crystals of dimension ~ 20.5 ± 4.2 µm appeared with $3CdSO_4.8H_2O$ as the only precipitant present. Over a large range of $C_{ammonium\_sulphate}$ ~ 0.02 – 2.0M, the crystals were observed to appear within ~24 hours and the extent of nucleation and the final size of crystals remained almost unaltered with crystal size attaining $l$ ~ 210 ± 5µm. These results signified vanishing effect of precipitant concentration on crystal nucleation on the $S_2|_{40\%}$ wrinkled surfaces. In another set of experiments, in which protein concentration was systematically varied, the crystal was observed to appear within 60 hours with protein concentration as low as 2 mg/ml on the $S_2|_{40\%}$ surface without use of any $(NH_4)_2SO_4$ but with 1.6% by weight of $3CdSO_4.8H_2O$ and 0.5mM $NaN_3$. No such crystal appeared on the S surface (Figure S2) at similar conditions. Importantly the crystal appeared not at the edge of the drop but away from it, towards the centre, implying that crystallization did not occur via evaporation induced super-saturation of the protein solution, but surface pattern induced decrease in the energy barrier for nucleation. Notice that the crystals appeared somewhat different when crystallized on different wrinkled surfaces at different concentrations of the precipitant. For example, the crystals on $S_2|_{40\%}$ surfaces, especially the one that appear in optical micrograph of figure 3(l) appeared dendritic. In order to confirm if these crystals indeed represented Ferritin, we carried out x-ray diffraction. The diffraction pattern, presented in figure S3, showed that the crystal was of protein and not of salt and they could be diffracted up to 6Å resolution. In fact, similar dendrite shape of ferritin crystal has been observed and analyzed by others as well[23], who have attributed the dendritic growth of ferritin to oligomers of different chain length present in the protein solution. In our experiment, the crystal was constrained to grow within a confined space between the two parallel plates, which too appear to have influence on the crystal shape.

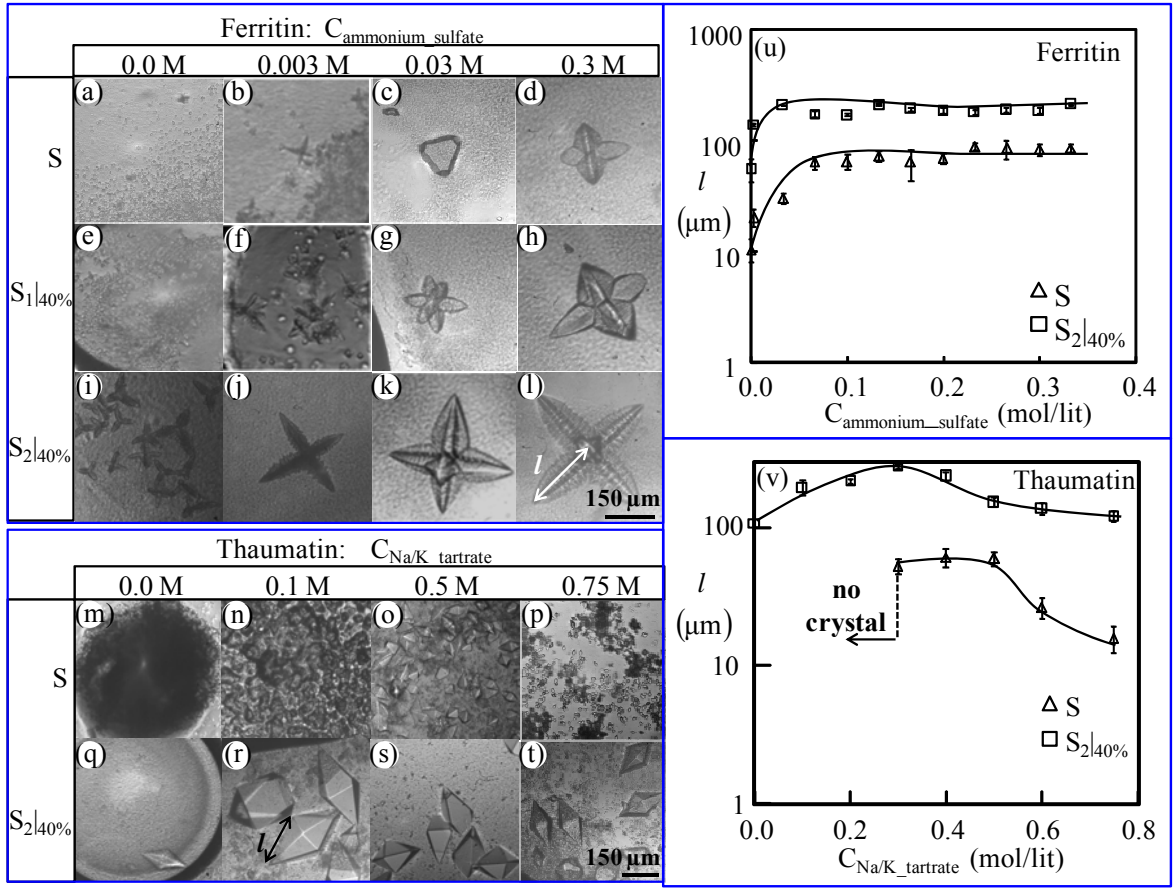

**Figure 3.** (a-l) The optical micrographs show ferritin crystals grown on three different substrates: S, $S_1$ and $S_2$ and at different concentration of ammonium sulfate, $C_{ammonium\_sulfate}$ = 0 to 0.3M as precipitant. Similarly, optical micrographs (m – t) represent tetragonal bipyramid shaped thaumatin crystals grown on two different substrates: S (m–p), $S_2|_{40\%}$ (q–t) by varying the precipitant concentration, $C_{Na/K\_tartrate}$ in the range 0 − 0.75 M. In figure (l), $l$ represents the longest dimension of crystal. Plot of $l$ as a function of precipitant concentration is plotted in figures (u) and (v) for ferritin and thaumatin respectively. The symbols Δ and □ represent substrates S and $S_2|_{40\%}$, respectively.

### 3.5. Crystallization of Thaumatin.

The above observations were made also with protein thaumatin (molecular weight = 22kDa), which was crystallized from a buffer solution in which sodium/potassium tartrate (Na/K tartrate) was used as the precipitant[24]. Figure 3m–t shows typical optical micrographs of crystals on S and

$S_2|_{40\%}$ surfaces with precipitant concentration, $C_{Na/K\_tartrate}$ varied over the range 0.0 – 0.75 M. On the control surface S, no crystal was observed to appear for $C_{Na/K\_tartrate}$ < 0.1M (Figure 3m–n); here the protein solution dried out completely over twelve hours. Beyond this threshold, the crystals appeared with their sizes increasing with the precipitant concentration. The crystal size maximized at $C_{Na/K\_tartrate}$ ~ 0.5 M (Figure 3o), beyond which, it decreased slightly; very small crystals appeared for $C_{Na/K\_tartrate}$ ~ 0.75 M as shown by micrograph in Figure 3p. The results were, however, remarkably different on the $S_2|_{40\%}$ surface (Figure 3q–t), for which the crystals of thaumatin appeared even without use of any precipitant (Figure 3q) unlike any example in the literature. In fact, the size of these crystals was already larger than those achieved on the S surface for various precipitant concentrations (Figure 3u–v). Thus, in effect, the $S_2|_{40\%}$ surface eliminated the need of a precipitant for this protein to crystallize. However, in order to produce even larger crystals, the $C_{Na/K\_tartrate}$ was optimized by varying it similarly, as on the S surface. Final result of these experiments was that crystals as large as 200.2 ± 19.6 µm, 155.6 ± 12.7 µm and 123.1 ± 9.3 µm appeared respectively for $C_{Na/K\_tartrate}$ ~ 0.1M, 0.5M and 0.75M (Figure 3r–t). These sizes far exceeded what could be achieved over the S surface, nevertheless were similar over three orders of magnitude in variation of $C_{Na/K\_tartrate}$. Similar to ferritin, for thaumatin too crystals appeared, albeit small, on the $S_2|_{40\%}$ surface even when the protein concentration was decreased by an order of magnitude to 1.5mg/ml with precipitant concentration as small as $C_{Na/K\_tartrate}$ = 0.2M. Here the crystals appeared within 48 hours. But the same did not happen on the featureless S surface (Figure S2), signifying once again the ability of the nano-wrinkled surface in crystallizing from a very dilute solution of protein and precipitant. It is worth noting that these concentrations of protein and the precipitant were significantly smaller than any reported value in the literature ~2 mg/ml)[15].

## 3.6. Crystallization of Glucose Isomerase and Xylanase II.

Glucose isomerase (molecular weight = 173 kDa) was crystallized from a solution containing 5mg/ml protein and 25 mM HEPES. In addition, $(NH_4)_2SO_4$ was used as the precipitant, the concentration of which was varied over $C_{ammonium\_sulphate} \sim 0.05 - 0.5$ M. On the control surface S, very small trigonal crystals appeared within 10 hours for $C_{ammonium\_sulphate} > 0.24$ M as depicted by the optical micrographs in Figure 4(a–c). With varying concentration of $(NH_4)_2SO_4$, the final size of the crystals was found to vary from $18.1 \pm 4.3$ μm to $33.0 \pm 4.8$ μm. However, on the $S_2|_{40\%}$ surface, crystals of size ~50 μm appeared within 24 hours at precipitant concentration as low as 0.05M; the size of the crystal grew to be as large as $l = 100.3 \pm 20.3$ μm for 0.5M $(NH_4)_2SO_4$ as shown in Figure 4(d–f), significantly larger than that achieved on the substrates S and on several other heterogeneous surfaces[25]. A typical x-ray diffraction pattern for a single glucose isomerase crystal grown on these nano-wrinkled substrates has been presented in Figure 4g for which maximum resolution of 2.5Å could be achieved. It is worth noting that similar crystal morphology could be observed earlier with significantly larger concentration of protein ~22 mg/ml on other commonly used heterogeneous substrates e.g. human hair[26].

It is to be noted that in all experiments described above, we used freshly prepared nano-wrinkled substrates. In order to see if these nano-wrinkled surfaces can be used also over a longer period, a typical $S_2|_{40\%}$ substrate was stored inside desiccators in dry and dust-free condition for three weeks, following which it was used for the crystallization experiment. The optical micrographs presented in figure 4(h) and (i) show the crystals which appear on these surfaces suggesting that ability of the $S_2|_{40\%}$ substrates to crystallize remain unaltered over this period, although nucleation gets somewhat uncontrolled.

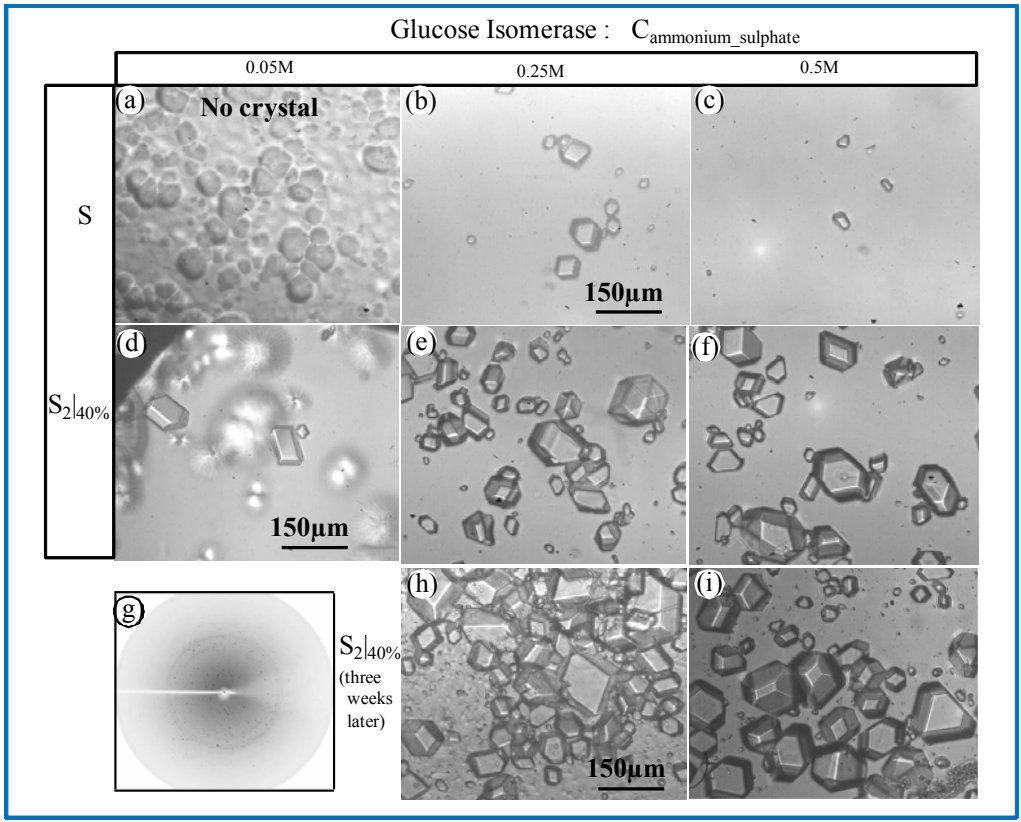

**Figure 4. Optical images (a−f) depict crystals of Glucose Isomerase grown by varying the precipitant concentration, $C_{ammonium\_sulphate}$ = 0.05 to 0.5 M on substrate S and $S_2|_{40\%}$. No crystal appeared of S surface for $C_{ammonium\_sulphate}$ < 0.25 M. (g) X-ray diffraction pattern from single glucose isomerase crystal. While in all the above cases, the PDMS surfaces were freshly prepared, optical images (h−i) show that crystals appear also on a $S_2|_{40\%}$ surface which was stored in a dust and moisture free environment for about three weeks. The concentration of Glucose Isomerase in the protein solution was maintained at 5 mg/ml for all these cases**

Similarly, protein xylanase II (molecular weight = 21kDa) too crystallized from the same initial concentration of 5 mg/ml. Plate like crystals were obtained from a solution containing 12.5 mM

NaCl, 6.25 mM Tris HCl, 1.8% (w/v) PEG 8000 in which, concentration of $CaCl_2$ was varied from 2.5 – 150 mM. Figure S4(b–c) depicts the optical micrograph of xylanase II crystals on the control surface S, in which large number of tiny crystals appeared corroborating the observation with the other protein molecules. With increasing concentration of precipitant, the crystal size increased to a certain extent on substrate $S_1|_{40\%}$ (optical micrographs in Figure S4(e–f)). However, the effect of precipitant became irrelevant on $S_2|_{40\%}$ on which, for all concentrations of the precipitant, only few crystals nucleated which grew to a significantly larger dimension ~ 208.4 ± 6.0 µm as presented in the optical micrographs in Figure S4(g–i) and Figure S5(b). The plate shaped crystals of Xylanase II were very thin with insignificant third dimension. In fact, they were so fragile that it was not possible to pick up the crystals from the mother liquor and to mount on the diffractometer for X-ray diffraction analysis. However, we could carry out dye test with Izit dye on these crystals, the optical micrographs presented in Figure S6 confirmed that these crystals indeed represented Xylanase II and not any salt. Furthermore, similar plate like morphology of these crystals has been observed to form earlier in experiments in which polysryrene nano-spheres are used as heterogeneous nucleant[27].

## 4. SUMMARY

The work presented here has several novel aspects which we will now summarize.

(a)     First of all, it is known that lengthscale of sub-microscopic features in soft materials is limited by their surface tension because surface tension does not allow very large curvature to be achieved in soft solids. Here we have overcome this limitation by introducing a simple two step process of preparing a substrate on PDMS. In fact, we have generated surface features which have curvature as large as 0.2 $nm^{-1}$. Furthermore, this method has allowed us to generate wrinkles, the mean curvature and the spread in distribution of curvature both can be varied over a large range.

(b)     Our theoretical analysis shows that both large curvature and large spread in curvature of a nucleant are useful for diminishing the energy barrier for nucleation of crystals. A substrate with such features is expected to trap protein molecules of wide range of sizes to crystallize and thus be a universal nucleant. We have shown that our nano-wrinkled surfaces comprising of folds with very large curvature, both positive and negative curvature and large distribution of curvature is useful for crystallizing protein molecules with sizes varying over a large range from 20 kDa to 480 kDa.

(c)     Experiments with protein molecules of wide range of molecular weight on identically prepared wrinkled surface, e.g. the $S_2|_{40\%}$ substrate, show that these molecules can indeed self-select a suitable lengthscale for heterogeneous nucleation which leads to crystallization over a wide range of protein and precipitant concentrations, not achieved with any other known substrate.

(d)     It is known that homogeneous nucleation occur only when concentration of the solute exceeds supersaturation, at which large number of crystals appear at a time, which do not grow to a very large size. However, for heterogeneous nucleation, crystal can appear at a much smaller extent of supersaturation[14,28]. In fact, supersaturation may not even be required in the bulk protein solution, as the nano-wrinkles having curvature similar to the radius of gyration of the protein molecule can lure these molecules resulting in protein concentration at the vicinity of the substrate significantly larger than that in the bulk. This effect is similar to the classical Kelvin effect observed in the context of capillary condensation, in which water is observed to condense inside a hydrophilic capillary even though the humidity in the atmosphere remains at sub-saturation level. Because of this effect, very few crystals appear which once nucleated, grow to a very large size thus consuming most of the solute molecules. As a result, further nucleation of crystal is prevented leading to highly controlled nucleation.

(e)    For all different protein molecules, crystals could be nucleated from protein solutions having wide range of precipitant concentration, importantly with very low concentration of precipitant. At least for one protein molecule, crystals appeared even without use of any precipitant. These results suggest that use of these nano-wrinkled surfaces is expected to simplify the screening process for crystallization of many yet-to-be-crystallized proteins, considered a bottleneck in molecular biology and its applications in therapeutics, disease diagnosis and its prevention.

(f)    By carrying out experiment on three week old $S_2|_{40\%}$ substrate, we have shown that the nano-wrinkled surfaces can be used for crystallization over a long period of time. More systematic experiments, however, needs to be carried out in order to understand the ageing effect of the substrates on their ability to crystallize protein molecules.

(g)    Beside protein crystallization, these surfaces are expected to have significant impact also in the general area of bio-mineralization, and variety of other operations involving nucleation, e.g. boiling, droplet condensation, protein separation and patterning.

**ACKNOWLEDGEMENTS**

Authors thank Miss Saloni Sharma, the staff of DST Unit on Soft Nanofabrication, IIT Kanpur for her help in carrying out the AFM scans and Dr. Krishnacharya Khare for useful discussions. ASG thank Mr. Vaibhav Singh Bais for his help in carrying out the X-ray diffraction analyses. ASG acknowledges Department of Science and Technology (DST), Government of India for the grant DST/CHE/20080204. AG acknowledges support from the Humboldt Foundation in the form of a fellowship for this work.

**ASSOCIATED CONTENT**

Images of different types of surface defects on nano-wrinkled substrates, images showing effect of nano-wrinkles on crystallizing thaumatin and ferritin from very low protein concentration in respective protein solutions, x-ray diffraction patterns of ferritin and glucose isomerase crystals, images depicting crystallization of Xylanase II on different surfaces using calcium chloride as the precipitant and in different precipitant concentrations, plot showing size of glucose isomerase and xylanase II crystals as a function of precipitant concentration on different surfaces, image showing plate shaped xylanase II crystal after dye test. This information is available free of charge via the Internet at http://pubs.acs.org/.

**For Table of Contents Use Only**

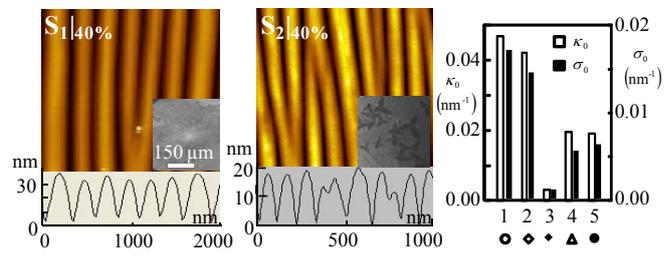

Table of Content Figure